# Upper-hybrid and electron-cyclotron waves in a laboratory magnetoplasma: weak spatial dispersion and parametric effects

Mikhail V. Starodubtsev and Alexander V. Kostrov

*Institute of Applied Physics, Russian academy of Science,
46 Ulyanov Street, 603950 Nizhny Novgorod, Russia*

Radiation of short-wavelength plasma waves is studied in a cold ($T_e = 0.5$ eV), large (1.2m length, 70 cm diameter), uniform, Maxwellian laboratory magnetoplasma in the upper-hybrid frequency range. Although the characteristic parameter of spatial dispersion $\beta^2 = V_{Te}^2/c^2$ is very small ($\beta^2 < 10^{-6}$), dispersion characteristics of the upper hybrid plasma waves is found to be strongly modified by the thermal motion of plasma electrons. Indeed, the resonance cone-like dispersion of UH plasma waves has been observed only if the parameter range satisfies simultaneously two inequalities: $\omega_c < \omega < 2\omega_c$, $\omega_p < \omega < \omega_{uh}$. In this parameter range, the maximal refractive index in the resonant directions has been found to be limited by $n_{max} \sim 15$; new UH mode with its dispersion determined by conical refraction has been found in the vicinity of the upper hybrid resonance ($\omega = \omega_{uh}$). Wave dispersion at $\omega > 2\omega_c$, $\omega_p < \omega < \omega_{uh}$ does not correspond to the cold plasma approximation and is determined by thermal plasma waves propagating along the ambient magnetic field.

Special attention is given to the wave propagation characteristics in the vicinity of the electron cyclotron resonance. An interesting effect of reduced electron cyclotron damping has been observed when two electromagnetic waves ($\omega_1 = \omega_c$, $\omega_2 = \omega_c + \Delta\omega$) propagate simultaneously with $\Delta\omega$ approximately equal to the lower hybrid frequency. This effect can be explained as a three-wave parametric process, analogous to the effect of electromagnetically induced transparency in three-level atomic system. Reduced electron cyclotron damping is extensively studied as a function of the beat frequency and the amplitude of the pump wave. Possible applications of the observed effect in plasma physics and electronics are discussed.

**Introduction**

Electrodynamic properties of magnetized plasmas in the so-called resonant frequency ranges characterize by significant growth of the refractive index *n*, when the angle $\varphi$ between the wave vector **k** and the ambient magnetic field **B**$_0$ approaches a critical value $\varphi_c$ [1]. In a cold plasma without energy losses, *n* tends to infinity at $\varphi_c$: $n(\varphi \to \varphi_c) \to \infty$, while all components of the dielectric permittivity tensor $\varepsilon(\omega, k)$ behave regularly and monotonically. Using analogy with usual frequency resonances, this effect could be interpreted as an *angular* resonance.

Resonant frequency ranges occupy two frequency domains (if plasma ions are immobile): $\omega_{LH} < \omega < \min\{\omega_p, \omega_c\}$ and $\max\{\omega_p, \omega_c\} < \omega < \omega_{UH}$, where $\omega_p$ and $\omega_c$ are the plasma and the electron cyclotron frequencies, $\omega_{LH}$ and $\omega_{UH}$ are the lower hybrid and the upper hybrid frequencies, respectively. We shall refer these frequency domains as the lower hybrid (LH) and the upper hybrid (UH) within the text. Typical wave index surfaces in resonant frequency ranges are shown in Fig.1 in the cold plasma approximation. Waves propagating along the resonant directions are quasi-electrostatic (**E** ∥ **k**) and characterize by large wave numbers *n*. Group velocities of these waves is orthogonal to their wave vector, as it is shown in Fig.1. Due to the plasma axial symmetry, group velocities form a resonance cone with opening angle $\theta_{res} = \pi/2 - \varphi_c$. When dissipation is neglected, an elementary dipole radiates only in an infinitely narrow vicinity of the resonance-cone surface. In real physical systems, however, there are always factors blurring the resonance; such factors include collisional and collisionless energy dissipation, finite emitter size, plasma nonuniformity and unsteadiness, finite width of the radiation frequency spectrum, etc. If all these factors are small enough, then the radiation is concentrated within a small solid angle near the resonance (group) cone. Along all other directions, the energy transfer, generally speaking, is not forbidden, but is suppressed by the angular resonance.

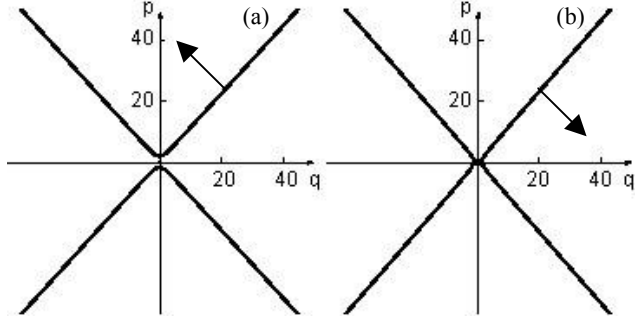

Fig.1. Wave index surfaces in LH (*a*) and UH (*b*) ranges. p = $k_z/k_0$; q = $k_\perp/k_0$

If finite plasma temperature is taken into account, dispersion of quasi-electrostatic plasma waves is modified due to the thermal motion of plasma electrons (i.e. by the plasma spatial dispersion). As result, refractive index does not tend to infinity at $\theta = \theta_{res}$, which blurs the angular resonance even if all other factors mentioned above are negligible. Moreover, new quasi-electrostatic plasma eigenmodes may occur.

Aim of this work is to study experimentally the influence of weak spatial dispersion on the dispersion properties of quasi-electrostatic waves in a laboratory magnetoplasma. The analysis is based on the examination of the precisely measured radiation patterns of a small electrostatically-coupled antenna.

**Experimental arrangement**

The experiment is performed in a large (radius 70 cm, length 1.5 m) laboratory plasma device. An axially magnetized ($B_0^{max}$ = 1000 G), uniform, Maxwellian afterglow plasma column with maximum density $n_e \approx 10^{13}$ cm$^{-3}$ is produced in argon ($p$ = 3 $10^{-3}$ Torr) by a pulsed inductive rf discharge (pulse duration 2 ms, repetition rate 0.2 Hz). Electron temperature $kT_e$ is equal to 5 eV immediately after finishing the discharge pulse and reaches $kT_e \approx kT_i \approx 0.3$ eV in late afterglow. Electron-ion collisions dominate in dense plasma ($\nu_e \approx \nu_{ei} \leq 10^8$ sec$^{-1}$) while in rare plasma (at $n_e \leq 3\ 10^9$ cm$^{-3}$) electron-neutral collisions are more important ($\nu_e \approx \nu_{em} \approx 2\ 10^5$ sec$^{-1}$). All measurements have been performed at low plasma densities $10^8$ cm$^{-3} \leq n_e \leq 10^{10}$ cm$^{-3}$, where electron collision rate is small ($\nu_e < 3\ 10^5$ sec$^{-1}$).

Wave characteristics have been studied in the center of the plasma column, where plasma density and the ambient magnetic field are highly uniform: $\Delta B_0/B_0 \leq 1\%$, $n_e/\nabla n_e \geq 5$ m. Short-wavelength quasi-electrostatic plasma eigenmodes are excited by a small insulated rf probe (length 10 mm, diameter $d$ = 1.5 mm). Maximal wave numbers of the radiation are limited by the probe size: $k_{max} \approx 1/d \approx 7$ cm$^{-1}$. Movable small noninsulated rf probe (length 10 mm, diameter 0.5 mm) is used to detect the radiation patterns of the emitter. The distance between the emitter and the receiver is 8 cm.

Radiation patterns of the emitter have been studied in a wide range of plasma parameters: $10^8$ cm$^{-3} < n_e < 10^{10}$ cm$^{-3}$, 30 G $< B_0 <$ 1000 G, at signal frequencies 100 MHz $< f = \omega/2\pi <$ 3 GHz. Using dimensionless parameters, we can represent the above parameter range as $0.05 < u < 100$, $0.01 < \upsilon < 100$, which naturally breaks into two resonant sub-ranges: the upper-hybrid (UH) with $0.05 < u < 1$, $1-u < \upsilon < 1$, and the lower-hybrid (LH) with $1 < u < 100$, $1 < \upsilon << n^2_{max}$, where $n_{max} = k_{max}c/\omega$ is the maximal refractive index produced by the emitting antenna. The latter inequality means that the emitter is substantially coupled to the quasi-electrostatic branch of the whistler mode.

The characteristic parameter of spatial dispersion $\beta^2_{Te} = (V_{Te}/c)^2$ ($V_{Te}$ and $c$ are the electron thermal velocity and velocity of light, respectively) was very small in the presented experiments: $\beta^2_{Te} = 10^{-6}$.

Damping processes did not influence the excitation of the resonance cones in this experiment. Thus the parameter of collisional damping, $\nu/\omega << 10^{-4}$, was too small to broaden the resonance. Smallness condition for Landau damping ($\omega >> k_{max}V_{Te}$, or $\beta^2_{Te} n_{max}^2 << 1$) has always been satisfied due to low plasma temperature: $10^{-4} < \beta^2_{Te} n_{max}^2 < 10^{-2}$. Finally, to avoid the electron cyclotron collisionless damping we will not consider the close vicinity of the electron gyrofrequency: all measurements have been done out of the 30-MHz slot around $f_c$ where the damping is significant (the smallness condition for the electron cyclotron damping is $\omega - \omega_c \gg k_{max} V_{Te} \approx 2\pi \times 30$ MHz).

Thus the measurements have been performed in a uniform, Maxwellian, collisionless, cold, afterglow laboratory magnetoplasma when propagation of short-wavelength plasma waves are not distorted by dissipative processes. Present experiments have purposed three main objectives: (i) to determine the parameter range where the dispersion properties of plasma waves conform to the cold plasma approximation, (ii) to find experimentally how the dispersion properties of UH waves modify due to the weak spatial dispersion and (iii) to compare the experimental results with existing theory of waves in a plasma with weak spatial dispersion.

**Experimental results**

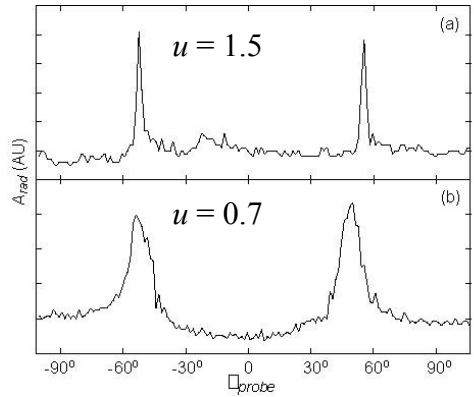

Fig.2. Resonance cones in LH (*a*) and UH (*b*) ranges.

Fig.2 demonstrate the comparison between resonance cones observed in the lower-hybrid (Fig.2a) and in the upper-hybrid (Fig.2b) domains. One can notice that in the LH domain, a small rf probe excites a very narrow resonance cone whose width is nearly equal to the probe diameter. This result is in a good agreement with the aboveestimations showing that all damping processes should not blur the resonance cones whose width is limited only by the probe size. UH resonance cones, on the contrary, are significantly wider than the emitter size, while all damping processes are still negligibly small and, hence, can not limit the Q-factor of the resonance.

The difference between these two cases is due to the plasma spatial dispersion, as it will be shown hereafter by theoretical analysis.

Resonance cone structures measured at different u are presented in Fig.3 and Fig.4 as a function of v. Fig.3 displays resonance cones in LH range ($u > 1$). Fig.3a shows the position of LH resonance cones calculated within the cold plasma approximation: Fig.3b-3f represent experimental results. We limited our studies by $u < 100$ in order to avoid the influence of ion movement on the wave properties. One may notice that the LH resonance cones behave in a fairly good agreement with the cold plasma approximation. Let us stress that within the parameter range represented in Fig.3, we may always observe well-formed resonance cones (at $1 < \upsilon < 10$) whose width is approximately equal to the emitter size. It means that plasma spatial dispersion does not modify dispersion characteristics of the quasi-electrostatic waves that form LH resonance cones.

Substantially different result are found when we examine the UH field patterns presented in Fig.4. Again, Fig.4a represent the calculated positions of UH resonance cones within the cold approximation; Fig.4b-4j display the laboratory measurements. One may observe that at *u* close to unity (Fig.4b-4c) the measured field patterns well correspond to the UH resonance cones in the cold plasma theory (let us only remind that the measured width of UH cones is noticeably greater than the expected value, see Fig.2). When *u* decreases (Fig.4d-4g), the measured field pattern start to deviate from the cold plasma approximation and a new plasma eigenmode occur at $u \approx \upsilon - 1$. This eigenmode is observed in Fig.4 as a narrow maximum along the ambient magnetic field; it is negligibly small at $u \approx 1$ (Fig.4b) but become much larger than the resonance cones at u ≈0.3 (Fig.4g). This eigenmode does not exist in the cold plasma approximation. At $\omega > 2\omega_c$, i.e. at u < 0.25 (Fig.4h and, especially, Fig.4i-4j), the field pattern produced by a small rf probe is determined mainly by this new mode, while the contribution of the resonance cones become negligible: Q-factor of the angular resonance become very low and the resonance cone structure can not be observed at any given $\upsilon$.

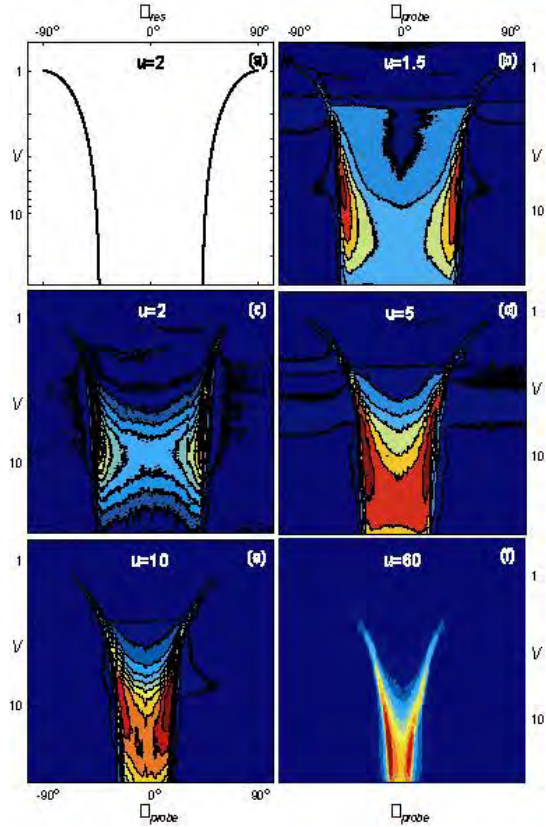

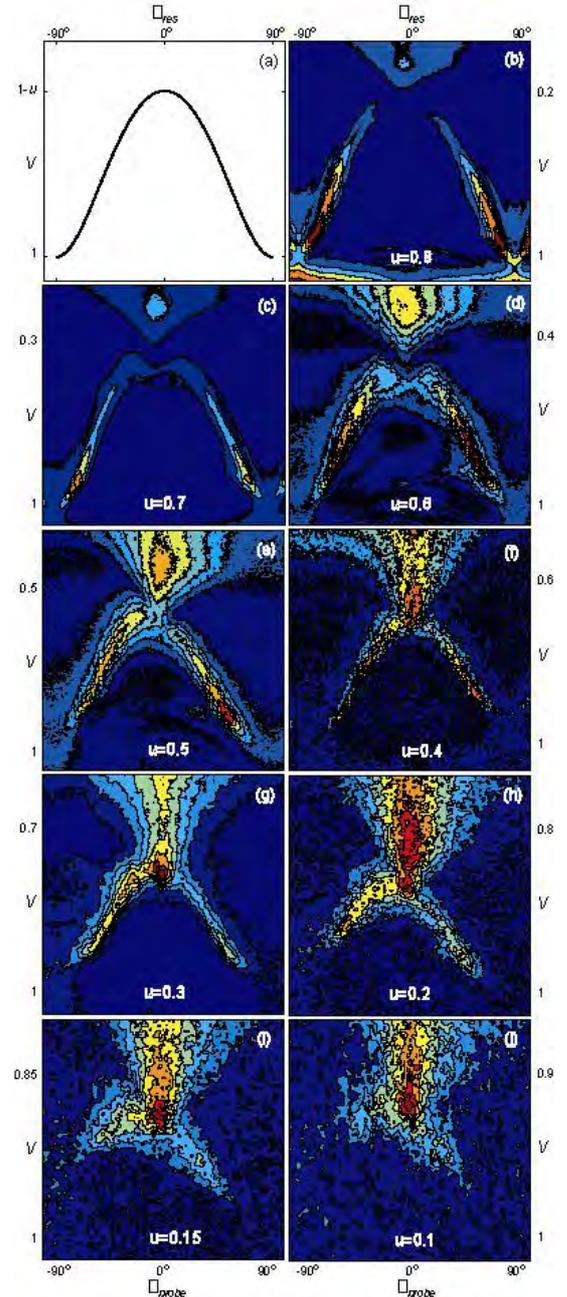

Fig.3*a* – position of LH resonance cones calculated within the cold plasma approximation; Fig.3*b-f* – Field patterns measured in LH frequency range.

Fig.4*a* – position of UH resonance cones calculated within the cold plasma approximation; Fig.4*b-j* – Field patterns measured in UH frequency range.

## Discussion

If energy dissipation is neglected, then the dispersion relation with weak thermal corrections takes the form [2]:

$$\delta n^6 + A n^4 + B n^2 + C = 0, \qquad (1)$$
$$\delta = -\beta^2_{Te}\, \upsilon\{3(1-u)\cos^4\theta + (6-3u+u^2)\sin^2\theta\cos^2\theta/(1-u)^2 + 3\sin^4\theta/(1-4u)\};$$
$A$, $B$ and $C$ are usual coefficient used in the cold plasma approximation. Dispersion relation describes three branches: ordinary waves (which we will not discuss), extraordinary waves, which form resonance cones in the cold approximation, and short-wavelength plasma waves. Two latter waves form one quasi-electrostatic branch, as it is shown in Fig.5. Let us point out that the angular resonance is strongly modified even at very small plasma temperature, when the effect of plasma spatial dispersion on other wave branches is negligible. Comparing Fig.1b and Fig.5, one may notice that the 'straight' part of the angular resonance is limited at $n^* \approx 20$, which coincides with the typical width of UH resonance cones. Let us also notice that the beamed radiation observed near the upper hybrid resonance ($\upsilon \approx 1 - u$) is not clear from the wave dispersion relation (1).

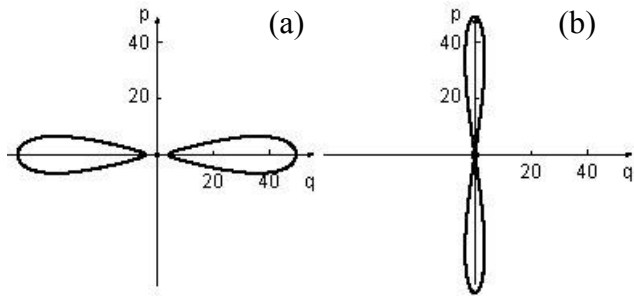

Fig.5. Wave index surfaces in UH range: (*a*) $u = 0.4$, $\upsilon = 0.615$; (*b*) $u = 0.4$, $\upsilon = 0.986$

**Conclusion**

Our laboratory experiments have evidenced the efficient excitation of plasma resonance cones. However, the parameter range within which the angular resonance is observed is significantly narrower than it follows from the analysis of the dispersion properties of a cold magnetized plasma. Main results of the experiment are as follows.
  (1) In the lower hybrid frequency range, the structure of resonance cones always is in a good agreement with the cold plasma approximation, i.e. the dispersion properties of quasi-electrostatic LH waves may be correctly described in the frame of this approximation.
  (2) The upper hybrid range is separated into two sub-ranges:
      (2a) $0.25 < u < 1$, $1 < \upsilon < 1 - u$ (i.e. $\omega_c < \omega < 2\omega_c$). Within this sub-range we have detected resonance cone-like structures, whose properties approximately agree with the cold plasma approximation, but the Q-factor of the angular resonance is smaller than it is predicted by the cold theory;
      (2b) $u > 0.25$, $1 < \upsilon < 1 - u$ (i.e. $\omega > 2\omega_c$). Within this sub-range no resonance cone structure is observed.
  (3) New plasma eigenmode propagating along the ambient magnetic field near the upper hybrid resonance ($\upsilon = 1 - u$) is observed.